\documentstyle[emulateapj,psfig]{article}

\def\be{\begin{equation}}
\def\ee{\end{equation}}
\def\Mesz{M\'esz\'aros}
\def\siml{\lower4pt \hbox{$\buildrel < \over \sim$}}
\def\simg{\lower4pt \hbox{$\buildrel > \over \sim$}}

\begin{document}

\title{Gamma-ray Burst Afterglow with Continuous Energy Injection:\\
Signature of a Highly-Magnetized Millisecond Pulsar}

\author{Bing Zhang \& Peter M\'esz\'aros}
\affil{Astronomy \& Astrophysics Dept.,
 Pennsylvania State University, University Park, PA 16802}

\begin{abstract} We investigate the consequences of a continuously
injecting central engine on the gamma-ray burst afterglow emission,
focusing more specifically on a highly-magnetized millisecond pulsar
engine.  For initial pulsar parameters within a certain region of the
parameter space, the afterglow lightcurves are predicted to show a
distinctive achromatic bump feature, the onset and duration of which
range from minutes to months, depending on the pulsar and the fireball
parameters. The detection of or upper limits on such features would
provide constraints on the burst progenitor and on magnetar-like
central engine models. An achromatic bump such as that in GRB 000301C
afterglow may be caused by a millisecond pulsar with $P_0=3.4$ms and
$B_p=2.7\times 10^{14}$ G.
\end{abstract}


\section{Introduction}

Much of the current research on Gamma-Ray Bursts (GRB) is aimed at
determining the nature of the central engine and its progenitor
system. While recently substantial results have begun to accumulate,
the evidence is still tentative.  Thus, the investigation of criteria
to differentiate between various central engine possibilities is
desirable.  Almost all the present fireball models, including those
considering various non-uniform injection scenarios, assume that the
energy injection into the fireball occurs in a short period of
time. This is also the case in the ``refreshed shock'' scenario (Rees
\& M\'esz\'aros 1998; Kumar \& Piran 2000; Sari \& \Mesz~ 2000).
However, in some types of central engines, such as a fast-rotating
high-field pulsar (magnetar) or a black hole plus a long-lived debris
torus system, a significant energy input into the fireball may in
principle continue for a time scale significantly longer than the
$\gamma$-ray emission. Therefore, there is a need to investigate a
continuously-fed fireball in more detail.  An additional motivation is
provided by the recent detection of Fe features in the X-ray afterglow
of GRB 991216 after about 1.5 days (Piro et al.  2000) and GRB 000214
after about 1 day (Antonelli et al, 2000), which may require a
continuing post-burst outflow in order to achieve less restrictive Fe
abundance constraints (Rees \& \Mesz, 2000).  Dai \& Lu (1998a,b)
first considered continuous injection from a millisecond pulsar to
interpret the afterglow light curves of some GRBs, but did not perform
a systematic study on this topic.  In this paper, we investigate the
observational consequences of a continuously injecting central engine,
and more specifically, focus on the possibility that the central
engine is a millisecond pulsar, in particular a highly magnetized
pulsar or magnetar.

\section{Continuous-injection dynamics}

We consider an engine which emits both an initial impulsive energy
input $E_{\rm imp}$ as well as a continuous luminosity, the latter
varying as a power law in the emission time.  In this case a
self-similar blast wave is expected to form at late times (Blandford
\& McKee, 1976).  The differential energy conservation relation for
the self-similar blast wave can be written as $dE/dt={\cal L}_0
(t/t_0)^{q'} -\kappa' (E/t)$, where $E$ and $t$ are the energy and
time measured in the fixed frame. The first term ${\cal L} ={\cal
L}_0(t/t_0)^{q'}$, where $q'$ and $\kappa'$ are dimensionless
constants, denotes the continuous luminosity injection, and the second
term takes into account radiative energy losses in the blast wave.
For $q'\neq-1-\kappa'$, an analytical solution is $E=\frac{{\cal
L}_0}{\kappa'+q'+1}\left (\frac{t}{t_0}\right)^{q'} t+ E_{\rm imp}
\left (\frac{t}{t_0}\right )^{-\kappa'}$ for $t>t_0$ (Cohen \& Piran
1999).  Here $t_0$ is a characteristic timescale for the formation of
a self-similar solution, which is roughly equal to the time for the
external shock to start to decelerate, and $E_{\rm imp}$ is a constant
which describes the impulsive energy input when $q'>-1-\kappa'$.  This
is obtained under the assumption that a self-similar solution exists
at $t>t_0$, hence it cannot be extrapolated down to $t=0$.  For
$t>t_0$ the bulk Lorentz factor of the fireball scales with time as
$\Gamma^2 \propto t^{-m}$, with $m$ and $\kappa'$ connected by
$\kappa'=m-3$ (Cohen, Piran \& Sari 1998), and $m=3$ for the adiabatic
case (Blandford \& McKee 1976).

In the observer frame the time $T$ is related to the fixed frame $t$
(for which $dr=c dt$) by $dT=(1-\beta)dt \simeq dt/2\Gamma^2$, and
$T=\int_0^{t}(2\Gamma^2)^{-1} dt\simeq {t}/[{2(m+1)\Gamma^2}]$ when
$t\gg t_0$.  The differential energy conservation relation can be now
expressed as $dE/dT=L_0 (T/T_0)^{q}-\kappa (E/T)$, and the integrated
relation is
\be
\begin{array}{ll}
E=\left(\frac{L_0}{\kappa+q+1}\right)\left(\frac{T}{T_{0}}\right)^q
T+E_{\rm imp}\left(\frac{T}{T_{0}}\right)^{-\kappa}, & T > T_0.
\label{E}
\end{array}
\ee
Here $L=L_0 (T/T_0)^{q}$ is the intrinsic luminosity of the central
engine, $T_{0}={t_0}/[{2(m'+1)\Gamma^2(t_0)}]$ where $m'$ is the
self-similar index for $t<t_0$ (usually $m'=0$ for the coasting
phase), $L_0=2\Gamma^2(t_0) {\cal L}_0$, $q=(q'-m)/(m+1)$, and
$\kappa=\kappa'/(m+1)$. Since $\kappa+q+1=(\kappa'+q'+1)/(m+1)$, the
comparisons between $q$ and $-1-\kappa$ in the following discussions
are equivalent to the comparisons between $q'$ and $-1-\kappa'$ (Cohen
\& Piran 1999).  Setting $T=T_0$, the total energy at the beginning of
the self-similar expansion is the sum of two terms, $E_0=L_0
T_0/(\kappa+q+1)+E_{\rm imp}$.  The first term, for $q>-1-\kappa$, is
the accumulated energy from the continuous injection (with radiative
corrections) before the self-similar solution starts (note that for
$q<-1-\kappa$ the two terms no longer have a clear physical meaning
since the first one is negative). The second term, $E_{\rm imp}$, is
the energy injected impulsively by the initial event.

At different times the total energy of the blast wave given by
Eq.(\ref{E}) may be dominated either by the continuous injection term
($\propto T^{(q+1)}$), or by the initial impulsive term ($\propto
T^{-\kappa})$.  Which of these two is dominant at a particular
observation time $T$ depends both on the relative values of the two
indices ($q+1$ and $-\kappa$, Cohen \& Piran 1999), {\em and} also on
the values of $L_0$ and $E_{\rm imp}$ (Dai \& Lu 1998a,b).  We then
have three regimes: i) If $q<-1-\kappa$, the second term in (\ref{E})
always dominates since the first term is negative.  The fireball is
then completely analogous to the impulsive injection case.  ii) If
$q=-1-\kappa$, the solution (\ref{E}) is no longer valid, and there is
no self-similar solution.  iii) If $q>-1-\kappa$, the first term in
(\ref{E}) will eventually dominate over the second term after a
critical time $T_c$, and it is this term that will exert a noticeable
influence on the GRB afterglow light curves. This latter case is of
the most interest here.

For a fireball blast wave decelerated by a homogeneous external medium
with particle number density $n$, the energy conservation equation at
time $t=r/c$ is
\be
\begin{array}{ll}
E=\frac{4\pi}{3} r^3 n m_p c^2 \Gamma^2=\frac{4\pi}{3} (ct)^3 n m_p
c^2 \Gamma^2, & t>t_0~,
\end{array}
\label{E''}
\ee
where $r$ is the radius of the blast wave, and all other symbols have
their usual meanings. This relation holds also for more general cases
with a non-constant $E$.
For the continuous injection dominated case, the energy $E$ in
(\ref{E''}) should have the same time dependence as the first term in
the RHS of Eq. (\ref{E}), giving $T^{q+1} \propto t^3\Gamma^2$, or
\be
m=\frac{2-q}{2+q},~~~~~(q>-1-\kappa).
\label{m-q}
\ee
Since in general $\Gamma\propto t^{-m/2} \propto r^{-m/2} \propto
T^{-m/2(m+1)}$ and $r\propto T^{1/(m+1)}$, using Eq. (\ref{m-q}) the
dynamics of the continuous injection-dominated case is
\be
\begin{array}{ll}
\Gamma\propto r^{-(2-q)/2(2+q)} \propto T^{-(2-q)/8}, &
                                              r\propto T^{(2+q)/4} ~.
\end{array}
\ee
For such a continuously-fed fireball, a forward shock propagating into
the external medium and a reverse shock propagating back into the
relativistic fireball will co-exist on either side of the contact
discontinuity. The latter may persist as long as a significant level
of energy injection is going on. The radiation spectra from these
shocks are complicated by many factors. Here we address only the
simplest case of the standard adiabatic external shock afterglow
scenario, and assume that the reverse shock is mildly relativistic, as
in the refreshed shock scenario. Following M\'esz\'aros \& Rees
(1997), Kumar \& Piran (2000) and Sari \& M\'esz\'aros (2000), one can
work out the relationship between the temporal index $\alpha$ and the
spectral index $\beta$, where $F_{\nu}
\propto T^{\alpha} \nu^{\beta}$. For the forward shock, the synchrotron
peak-frequency $\nu_m^f \propto \Gamma B' \gamma_m^2\propto \Gamma^4
\propto T^{-2m/(m+1)}\propto T^{-(2-q)/2}$, and the peak flux 
$F_{\nu_{m}}^f \propto (t^2\Gamma^5) (n_e' B' r/\Gamma) \propto T^3
\Gamma^8 \propto T^{(3-m)/(1+m)} \propto T^{1+q}$, so that
$\alpha^f=(2m\beta^f+3-m)/(1+m)=(1-q/2)\beta^f+1+q$. For the reverse
shock, $\nu_m^r=\nu_m^f/\Gamma^2\propto \Gamma^2
\propto T^{-m/(m+1)}\propto T^{-(2-q)/4}$, and $F_{\nu_{m}}^r =
\Gamma F_{\nu_{m}}^f \propto T^3 \Gamma^9 \propto$ $T^{(6-3m)/2(1+m)}
\propto T^{(6+9q)/8}$, so that $\alpha^r=(2m\beta^r+6-3m)/2(1+m)=
[(4-2q)\beta^r+9q+6]/8$. At any time the emission at a given
frequency, may be dominated either by the forward or the reverse
shock. The above scalings are derived for the slow cooling regime
(Sari, Piran \& Narayan 1998), which is usually satisfied. The fast
cooling will change the scaling laws, but does not change the
qualitative ``index switching'' picture proposed in this paper. All
the above scalings reduce to the standard adiabatic case by setting
$m=3$ and $q=-1$.

The injection-dominated regime begins at a critical time $T_c$ defined
by equating the injection and energy loss terms in (\ref{E}),
\be
T_{c}={\rm Max}\left\{1,\left[(\kappa+q+1)
\left(\frac{E_{\rm imp}}{L_{0}T_{0}}\right)\right]
^{1/(\kappa+q+1)} \right\}T_{0},
\label{tec}
\ee
where $T_{c}\geq T_{0}$ ensures that a self-similar solution has
already formed when the continuous injection law dominates. If
initially the continuous injection term is more important, i.e.,
$L_{0}T_{0} \ge E_{\rm imp}$, then $T_c \simeq T_{0}$, and the
dynamics is determined by the continuous injection law as soon as the
self-similar profile forms.  However, if initially the impulsive term
dominates ($L_{0}T_{0}\ll E_{\rm imp}$), the critical time $T_c$ after
which the continuous injection becomes dominant could be much longer
than $T_{0}$, depending on the ratio of $E_{\rm imp}$ and
$L_{0}T_{0}$.

Central engines with a continuous injection may, in addition, have
another characteristic timescale ${\cal T}$, e.g. at which the
continuous injection power law index (say $q_1 >-1-\kappa$) switches
to a lower value $q_2 <-1-\kappa$. It is only for ${\cal T}>T_{c}$
that the continuous injection has a noticeable effect on the afterglow
light curve. Different central engines may have different values of
${\cal T}$.  In the following, we investigate the conditions under
which a continuous injection can influence the dynamics of the blast
wave, and we consider the specific case of a millisecond pulsar as a
central engine with a continuous injection following after the initial
impulsive phase.

\section{Highly magnetized millisecond  pulsar as the central engine}

Although a wide range of GRB progenitors lead to a black hole - debris
torus system (Narayan, Paczy\'nski \& Piran 1992; Woosley 1993;
Paczy\'nski 1998; M\'esz\'aros, Rees \& Wijers 1999; Fryer, Woosley \&
Hartmann 1999), some progenitors may lead to a highly-magnetized
rapidly rotating pulsar (e.g. Usov 1992; Duncan \& Thompson 1992;
Thompson 1994; Yi \& Blackman 1998; Blackman \& Yi 1998; Klu\'zniak \&
Ruderman 1998; Nakamura 1998; Spruit 1999; Wheeler et al. 2000;
Ruderman, Tao \& Klu\'zniak 2000). During the early stages of the
evolution of these objects, the luminosity decay law could be very
complicated.  On the longer (afterglow) timescales that we are
interested in, some short-term processes, such as the decay of the
differential-rotation-induced toroidal magnetic field energy
(Klu\'zniak \& Ruderman 1998; Ruderman et al. 2000), are no longer
important, and the energy injection into the fireball may be mainly
through electromagnetic dipolar emission. The spindown of the pulsar
may also be influenced by gravitational radiation.

Assuming that the spindown is mainly due to electromagnetic (EM)
dipolar radiation and to gravitational wave (GW) radiation, the
spindown law is $-I\Omega\dot\Omega= {(B_p^2R^6\Omega^4)}/{(6c^3)} +
{(32GI^2\epsilon^2\Omega^6)}/{(5c^5)}$ (Shapiro \& Teukolsky 1983),
where $\Omega$ and $\dot\Omega$ are the angular frequency and its time
derivative, $B_p$ is the dipolar field strength at the poles, $I$ is
moment of inertia, $R$ is stellar radius, $\epsilon$ is ellipticity of
the neutron star.  This equation can be solved for $\Omega$ as a
function of $T$, with initial conditions $\Omega =\Omega_0$,
$\dot\Omega=\dot\Omega_0$ for $T=0$.  The decay solution $\Omega(T)$
includes both EM and GW losses, but the corresponding energy input
into the fireball will be due to the EM dipolar emission only,
i.e. $L(T)={[B_p^2R^6\Omega(T)^4]}/ {(6c^3)}$, which usually will not
be a simple power-law. However, at different times the spindown will
be dominated by one or the other loss terms, and one can get
approximate solutions. When EM dipolar radiation losses dominate the
spindown, we have $\Omega=\Omega_0 (1+T/{\cal T}_{em})^{-1/2}$, or
approximately $\Omega= \Omega_0$ for $T\ll {\cal T}_{em}$, and
$\Omega=\Omega_0(T/{\cal T}_{em})^{-1/2}$ for $T\gg {\cal
T}_{em}$. Here
\be
{\cal T}_{em}=\frac{3c^3I}{B_p^2R^6\Omega_0^2} =2.05\times10^3~{\rm
s}~ I_{45}B_{p,15}^{-2}P_{0,-3}^2R_6^{-6},
\label{Teem}
\ee
is the characteristic time scale for dipolar spindown,
$B_{p,15}=B_p/(10^{15}{\rm G})$, and $P_{0,-3}$ is the initial
rotation period in milliseconds.  When GW radiation losses dominate
the spindown, the evolution is $\Omega \simeq \Omega_0{(1+T/{\cal
T}_{gw})^{-1/4}}$ where ${\cal
T}_{gw}=({5c^5}/{128GI\epsilon^2\Omega_0^4})=$ $0.91~{\rm s}~
I_{45}^{-1}P_{0,-3}^{4}(\epsilon/0.1)^{-2}$.  GW spindown is important
only when the neutron star is born with an initial $\Omega_0 \simg
\Omega_{\ast} \sim 10^4 {\rm s}^{-1}$ (e.g. Usov 1992; Blackman \& Yi
1998).  In such cases the ellipticity is large ($\epsilon\sim 0.1$)
due to rotational instability, and the timescale for the GW-dominated
regime is short, so that $\Omega$ will be damped to below
$\Omega_{\ast}$ promptly in a time ${\cal T}_\ast=
[({\Omega_0}/{\Omega_\ast})^4-1]{\cal T}_{gw}$. After $\Omega
<\Omega_{\ast}$, GW losses decrease sharply, and the spindown becomes
dominated by the EM losses. If the neutron star is born with $\Omega_0
< \Omega_{\ast}$, the spindown will be always in the EM regime, since
the typical spindown time for GW radiation is much longer.

For these regimes, the continuous injection is:\\ (A)
$\Omega_0<\Omega_{\ast}$, EM loss dominated regime. In this case
\be
\begin{array}{ll}
L(T)&=L_{em,0}\frac{1}{(1+T/{\cal T}_{em})^2}\\ & \simeq \left\{
\begin{array}{ll}
L_{em,0}, & T\ll {\cal T}_{em} \\ L_{em,0}\left(\frac{T}{{\cal
T}_{em}}\right)^{-2}, & T\gg {\cal T}_{em},
\end{array}
\right.
\end{array}
\label{Leem}
\ee
where ${\cal T}_{em}$ is given by (\ref{Teem}), and
\be
L_{em,0}=\frac{I\Omega_0^2}{2{\cal T}_{em}}\simeq 1.0\times10^{49}
{\rm erg~s^{-1}}B_{p,15}^2P_{0,-3}^{-4}R_6^6.
\label{Leem0}
\ee
\\
(B) $\Omega_0>\Omega_{\ast}$: initially GW dominated regime. The
continuous injection luminosity can be divided into two phases, i.e.,
$L = L_{em,0}/({1+T/{\cal T}_{gw}})$ for $T<{\cal T}_\ast$ or $L=
L_{em,\ast}/[{1+(T-{\cal T}_\ast)/{\cal T}_{em,\ast}}]^2$ for $T>{\cal
T}_\ast$, where $L_{em,0}$ is given by (\ref{Leem0}), and
$L_{em,\ast}={I\Omega_{\ast}^2}/{2{\cal T}_{em,\ast}}\simeq 1.0
\times10^{49}
{\rm erg~s^{-1}}B_{p,15}^2 P_{\ast,-3}^{-4}R_6^6$, where ${\cal
T}_{em,\ast}={3c^3I}/{B_p^2R^6\Omega_\ast^2}\simeq 2.05\times
10^3~{\rm s}~ I_{45}B_{p,15}^{-2}P_{\ast,-3}^{2}R_6^{-6}$.

The above continuous injection luminosities have, for certain times, a
temporal index $q=0>-1$, which may dominate the blast wave dynamics.
The typical duration times for this flat injection law are ${\cal
T}_{em}$ for case (A) or ${\cal T}_{gw}$ and ${\cal T}_{em,\ast}$ for
case (B). In case (B) there are two luminosity ``plateaus''. The
former is usually much shorter than $T_c$ unless a very dense medium
is assumed (see below), so it is unlikely to detect such a
``two-step'' injection-dominated case.  If $\Omega_\ast$ is close to
$\Omega_0$, the second timescale for case (B), ${\cal T}_{em,\ast}$,
may not be much different from ${\cal T}_{em}$. In the following, for
simplicity, we discuss case (A) only, keeping in mind the possible
extra complexity which case (B) may introduce in some extreme cases.

During the time interval $T_{c}<T<{\cal T}_{em}$ one can expect a
distinctive pulsar feature to show up in the lightcurve.  We take as
an example the behavior (\ref{Leem}), setting $q=0$ in this regime,
and $q=-1$ otherwise (since the second slope $q=-2$ would mimic in
effect the standard impulsive $q=-1$, $m=3$ adiabatic case).
For the slow-cooling external shock scenario (\S 2), the temporal
decay index changes at $T_c$ from $\alpha_1$ to
$\alpha_2=(2/3)\alpha_1+1$, and changes back to $\alpha_1$ after
${\cal T}_{em}$. The temporal index is related to the spectral slope
$\beta$ through $\alpha_1=(3/2)\beta$ for the forward shock, and
$\alpha_1= (6\beta-3)/8$ for the reverse shock.  This implies an {\em
achromatic} bump in the light curve, which could provide a signature
for a pulsar.  However, not all millisecond pulsars would give such
observable signatures.  The condition for detecting this feature is
${\cal T}_{em}>T_{c}$, which constrains the pulsar initial parameter
phase space.  Let us specify the pulsar case in Eq.(\ref{tec}), i.e.,
$q=0$, and further assume $\kappa=0$, so that Eq.(\ref{tec}) is
simplified to $T_c= {\rm Max}(1,E_{\rm imp}/L_0 T_0)T_0$. This gives
two possibilities:

(I) $E_0$ is mainly due to the continuous injection, i.e., $E_{\rm
imp} \lesssim L_0T_0$. This is the case considered in many pulsar GRB
central engine models (Usov 1992; Duncan
\& Thompson 1992; Thompson 1994; Spruit 1999).
We then have $T_c=T_0$ and $E_0\simeq L_{em,0}T_0$ (a factor of 2 if
the continuous injection and the prompt injection energies are
comparable). Solving $E_0=(4\pi/3)(2c\Gamma_0^2 T_0)^3nm_p c^2
\Gamma_0^2$ (in which $m'=0$ has been adopted), where
$\Gamma_{0}=\Gamma(T_0)=E_0/\Delta M c^2$ is the initial bulk Lorentz
factor of the blast wave, we have $T_c=T_0 \simeq 0.33{\rm s}~
B_{p,15} P_{0,-3}^{-2}R_6^3 (\Gamma_{0}/300)^{-4} n^{-1/2}$.  The
condition ${\cal T}_{em}>T_0$ then implies
\be
B_{p,15}<18.4 P_{0,-3}^{4/3} I_{45}^{1/3} R_6^{-3}
(\Gamma_{0}/300)^{4/3}n^{1/6}.
\label{1}
\ee

(II) $E_0$ is dominated by the impulsive term $E_{\rm imp}$. This
could be the case if the central engine is a pulsar, the initial
impulsive GRB fireball being due, e.g. to $\nu{\bar\nu}$ annihilation
(e.g. Eichler et al 1989), dissipation of initial differential
rotation (Ruderman et al. 2000), phase conversion from a neutron star
to a strange star (Dai \& Lu 1998b), etc.  In this case, since $E_{\rm
imp}> L_0T_0$, we have $T_c=E_{\rm imp}/L_{0,em}= (2E_{\rm
imp}/I\Omega_0^2){\cal T}_{em}$. The condition ${\cal T}_{em} >T_c$ is
simply $E_{\rm imp}<(1/2)I\Omega_0^2$, or
\be
P_{0,-3}<4.4 I_{45}^{1/2}E_{\rm imp,51}^{-1/2}.
\label{2}
\ee
Since both ${\cal T}_{em}$ and $T_c$ are large in this case, the
continuous injection term may dominate at a later time.  An additional
constraint, to avoid rotational break-up of the pulsar, is
\be
P_0 > P_{0}({\rm min})~.
\label{3}
\ee
The equations (\ref{1}), (\ref{2}) and (\ref{3}) define a region of
the pulsar $P_0,B_{p,0}$ initial parameter space (see Fig. 1), inside
which the continuous injection has the observational signature
discussed above, i.e. a bump in the light curve. The separation
between the two regimes (I,II) is defined by the condition
$L_{em,0}T_0 \simeq 10^{51} E_{\rm imp,51}$ erg, or $B_{p,15} \simeq
6.7 P_{0,-3}^2 R_6^{-3}(\Gamma_{0}/300)^{4/3} n^{1/6} E_{\rm
imp,51}^{1/3}~$.  Above this line (regime I), $T_c \sim T_0$ and one
expects only one change, starting with the flat regime and changing to
steep; below this line (regime II) one expects two changes, starting
with the steep decay, followed by a flat regime, and a final
resumption of the steep decay.

\centerline{}
\centerline{\psfig{file=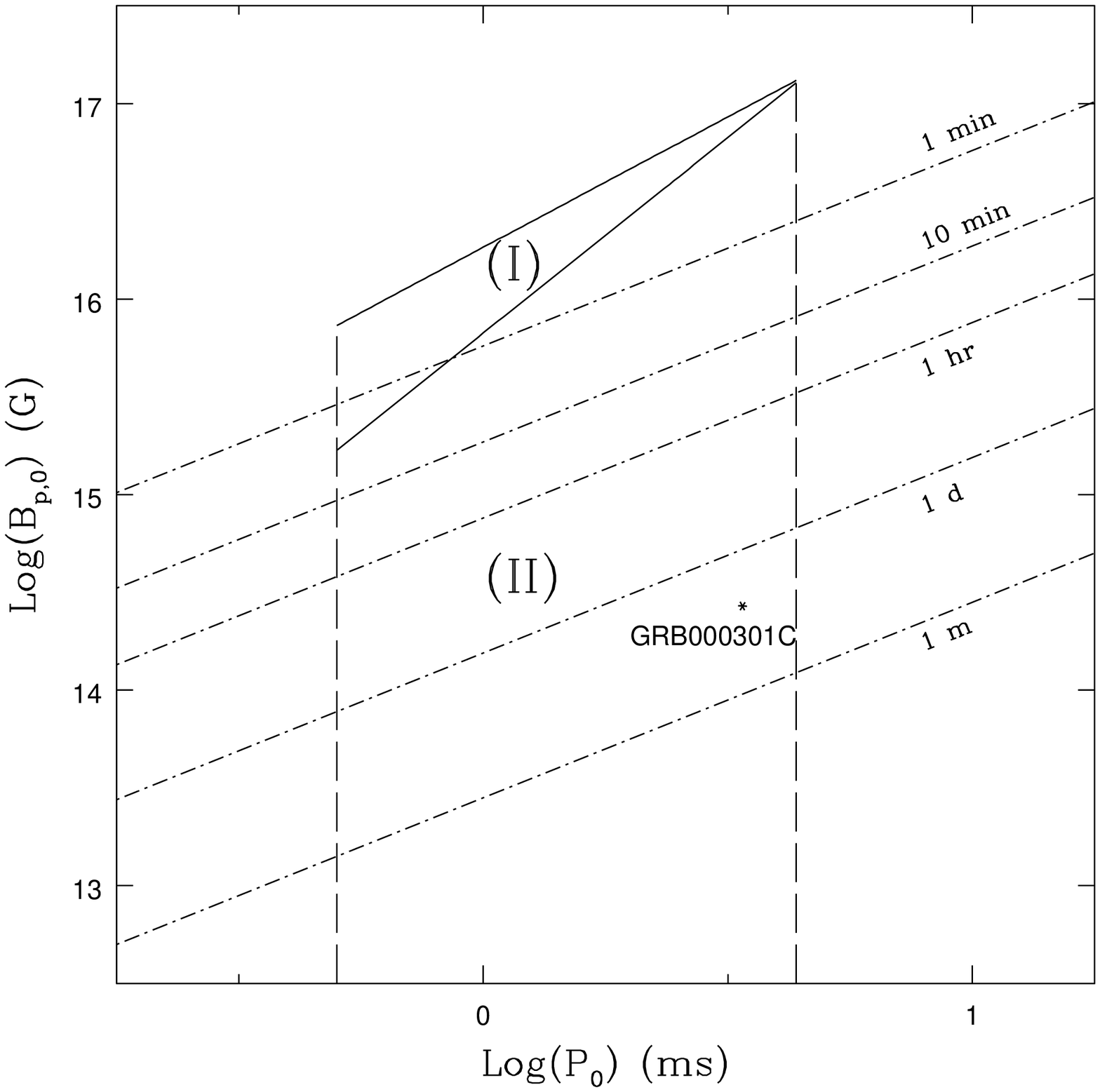,width=8.5cm}}
\figcaption{A $B_p-P_0$ diagram for the initial parameters of a pulsar
born in a GRB. The enclosed areas (regime I and II) are the phase
spaces where an achromatic pulsar signature is expected in the GRB
afterglow lightcurves. Dotted-dashed lines denote various ${\cal T}_{em}$
expected. $E_{\rm imp}=10^{51} {\rm erg~s^ {-1}}$, $\Gamma_{0}=300$,
$n=1 {\rm cm}^{-3}$, and $P_0({\rm min})=0.5$ ms have been adopted.}
\centerline{}

\section{Discussion}

If the central engine of GRB are fast-rotating pulsars, the afterglow
lightcurves may show a distinctive, achromatic feature, for pulsars
whose initial parameters are within ``pulsar signature region'' region
defined by equations (\ref{1}), (\ref{2}) and (\ref{3}). In this case
the afterglow light curves flatten after a critical time $T_c$
(eq.[\ref{tec}]), and steepen again after a time ${\cal T}_{em}$
(eq.[\ref{Teem}]).  This region of pulsar parameter phase space
includes ``magnetars" or ultra-high field pulsars, which play a large
role in some GRB pulsar progenitor models.

The current data on early afterglows are insufficient to provide good
tests for this feature. However, an interesting possibility is the
achromatic bump observed in the afterglow of GRB 000301C, which for a
limited time deviates significantly from the standard broken power-law
fit (Masetti et al. 2000).  Possible explanations include running into
a non-uniform ambient density (Berger et al. 2000), and modification
by a microlensing event (Garnavich, Loeb \& Stanek 2000). We suggest
here a third possibility, that the bump may be caused by the pulsar
signature discussed above. Taking (Berger et al 2000) $\alpha_1
\sim-1.28$ for the principal temporal index (before the bump and before
the decay ascribed to a jet transition), the temporal index during the
pulsar signature is expected to be $\alpha_2\sim 0.15$, which seems
reasonable to fit the achromatic bump. Since the feature occurs a
couple of days after the GRB trigger, the initial pulsar parameters
are in regime II. Taking $E_{\rm imp,51}\sim 1.1$ from the
observations (Berger et al. 2000), the pulsar parameters follow from
setting ${\cal T}_{em}\sim3.8$ d and $T_c\sim2.5$ d, which gives
$P_0\sim 3.4$ ms and $B_{p,15}\sim 0.27$ (Fig.1). Detailed $\chi^2$
fit to the data would be necessary to validate this proposal.

Another relevant observation may be the recent Fe line detection in
the X-ray afterglow of GRB 991216 (Piro et al 2000), which in one
interpretation would require a continuously-injecting central engine
(Rees \& M\'esz\'aros 2000). If $L_{em}\sim 10^{47} {\rm erg~s^ {-1}}$
is assumed 37 hours after the burst (Rees \& M\'esz\'aros 2000), this
continued injection could be due to a pulsar with $B_{p,15}\sim 0.15$
and $P_{0}<1.2$ ms, which would imply a signature bump at $\simg 1$
hour.  Such a bump is not seen in this afterglow (e.g. Halpern et
al. 2000).  However, a slightly weaker luminosity (e.g. $L_{em} \leq
3\times 10^{46} {\rm erg~s^ {-1}}$) could also explain the Fe line
features, by assuming a slightly larger Fe abundance (which is very
low in this model). The required pulsar can therefore be more
magnetized, and the characteristic times for the signature bump would
be expected early enough to have evaded detection in this GRB.

In conclusion, the detection of or upper limits on such characteristic
afterglow bumps by missions such as {\em HETE2} or {\em Swift} may be
able to provide interesting constraints on magnetar GRB models and
their progenitors.

\acknowledgments{We thank Z.G. Dai, M.J. Rees and a referee for comments,
and NASA NAG5-9192, NAG5-9193 for support.}

\end{document}